\documentclass[mathleft]{an}
\usepackage{graphicx}
\usepackage{color}
\usepackage{times}
\usepackage{amssymb}
\begin{document}

% The following seven commands are intended for editorial usage and should be
% ignored by the author(s).
\Pagespan{743}{}% Document's page range. 
% If second parameter is left empty, the last page is computed automatically.
\Yearpublication{2007}%
\Yearsubmission{2007}%
\Month{10}%   
\Volume{328}%  
\Issue{8}% 
 \DOI{10.1002/asna.200710795}% 
% \DOI{This.is/not.aDOI}% 

\title{Numerical simulations of fast and slow coronal mass ejections}

\author{T. T\"or\"ok\inst{1}\fnmsep\thanks{Corresponding author:
\email{tt@mssl.ucl.ac.uk}\newline}
\and B. Kliem\inst{2,3}
}
\titlerunning{Numerical simulations of fast and slow CMEs}
\authorrunning{T. T\"or\"ok \& B. Kliem}
\institute{University College London, Mullard Space Science Laboratory,
Holmbury St. Mary, Dorking, Surrey, RH5 6NT, U.K.
\and 
Astrophysical Institute Potsdam, An der Sternwarte 16, 
D-14482 Potsdam, Germany
\and
Kiepenheuer-Institut f\"ur Sonnenphysik, Sch\"oneckstra{\ss}e 6,
                  D-79104 Freiburg, Germany}

\received{2007 Mar 26}
\accepted{2007 May 24}
\publonline{2007 Sep 19}


\abstract{Solar coronal mass ejections (CMEs) show a large variety in their
kinematic properties. CMEs originating in active regions and accompanied 
by strong flares are usually faster and accelerated more impulsively than 
CMEs associated with filament eruptions outside active regions and weak 
flares. It has been proposed more than two decades ago that there are two 
separate types of CMEs, fast (impulsive) CMEs and slow (gradual) CMEs. 
However, this concept may not be valid, 
since the large data sets acquired in recent years do not show two distinct 
peaks in the CME velocity distribution and reveal that both fast and slow 
CMEs can be accompanied by both weak and strong flares. We present numerical 
simulations which confirm our earlier analytical result that a flux-rope
CME model
permits describing fast and slow CMEs in a unified manner. We consider a 
force-free coronal magnetic flux rope embedded in the potential field of 
model bipolar and quadrupolar active regions. The eruption is driven by 
the torus instability which occurs if the field overlying the flux rope 
decreases sufficiently rapidly with height. The acceleration profile depends 
on the steepness of this field decrease, corresponding to fast CMEs for rapid 
decrease, as is typical of active regions, and to slow CMEs for gentle
decrease, as is typical of the quiet Sun. Complex (quadrupolar) active
regions lead to the fastest CMEs.}

\maketitle

%=======================================================================
\section{Introduction}
%=======================================================================
Coronal mass ejections (CMEs) are spontaneous ejections of plasma and 
magnetic flux from the inner solar corona into interplanetary space. 
They can originate in active regions, where they are associated with
X-ray or EUV flares and often also with filament eruptions, but they can
also originate from eruptions of large quiescent filaments 
outside active regions not accompanied by strong flares. Early 
observations, based on relatively small data sets, indicated that 
flare-related CMEs are faster, often moving already at nearly constant 
speed (mostly in excess of $\approx750$~km\,s$^{-1}$)
when they enter the field of view of white-light coronagraphs, i.e.,
they are fully accelerated at heliocentric distances $< 2 R_{\odot}$,
whereas CMEs with weak or no flare activity are accelerated only
gradually to moderate speeds (typically $\approx400\mbox{--}600$~km\,s$^{-1}$)
within the coronagraph field of view of 6~$R_{\odot}$ or more.   

Such data led MacQueen \& Fisher (1983) to suggest the existence 
of two separate types of CMEs, flare-associated ones and
eruption-asso\-cia\-ted ones (Fig.\,1), the latter referring to
quiescent filament eruptions.
Their concept has been supported by Sheeley et al. (1999), who categorised
the two types as impulsive and gradual, respectively. Both author groups
suggested that the acceleration is dominated by different physical
processes, which may imply the need for different models.

Studies of the large CME data sets meanwhile available revealed a very
wide range of kinematic properties, with velocities ranging from 
$\sim\!50\,\mbox{--}\,3000$~km\,s$^{-1}$ and accelerations ranging from 
$\sim\!\!10$ to $\!>\!\!10^4$~m\,s$^{-2}$ over
$\sim 5\,\mbox{--}1000$ minutes. However, the data always showed a continuous 
distribution of velocities with a single peak (Zhang \& Dere 2006 and
references therein). Furthermore, both fast and slow CMEs can be accompanied 
by both weak and strong flares (Vr{\v s}nak, Sudar \& Ru{\v z}djak 2005). 
These findings suggest that the concept of physically different CME types 
might not be valid, rather there may exist a continuous distribution of 
CME kinematics, with fast impulsive and slow gradual CMEs representing
the two ends of the distribution.

%\begin{figure*}
%\begin{minipage}[t]{0.68\linewidth} 
%\includegraphics[width=0.49\linewidth]{toeroek_fig1a.ps}
%\hspace{1mm}
%\includegraphics[width=0.49\linewidth]{toeroek_fig1b.ps}
%\end{minipage}\hfill
%\begin{minipage}[b]{0.30\linewidth} 
%\caption{
%{\em Left:} observation of two apparently distinct CME types (data 
%from MacQueen \& Fisher 1983): fast CMEs (red) acquire nearly constant 
%speeds low in the corona (behind the coronagraph's occulting disk indicated 
%in grey), while slow CMEs (blue) are accelerated only gradually to reach the 
%gravitational escape speed (dashed line). {\em Right:} the TI of a freely expanding 
%current ring yields similar velocity profiles for different decay indices $n$ of the 
%external poloidal field $B_\mathrm{ex}$ (equilibrium field strengths at
%the initial radius $R_0\!=\!10$~Mm are given).\label{fig1}}
%\end{minipage}
%\end{figure*}

\begin{figure*}
\begin{minipage}[t]{0.68\linewidth}  %70
\includegraphics[width=1.0\linewidth]{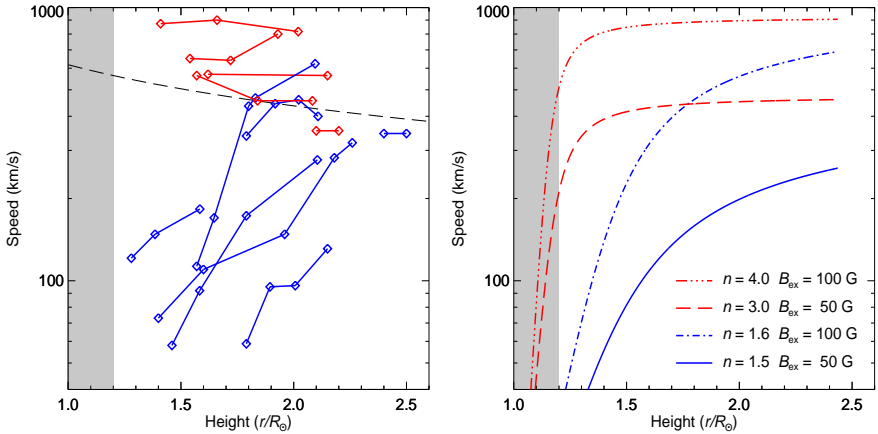}
%\hspace{1mm}
%\includegraphics[width=0.49\linewidth]{toeroek_fig1b.ps}
\end{minipage}\hfill
\begin{minipage}[b]{0.30\linewidth}   %28
\caption{
{\em Left:} observation of two apparently distinct CME types (data 
from MacQueen \& Fisher 1983): fast CMEs (red) acquire nearly constant 
speeds low in the corona (behind the coronagraph's occulting disk indicated 
in grey), while slow CMEs (blue) are accelerated only gradually to reach the 
gravitational escape speed (dashed line). {\em Right:} the TI of a freely expanding 
current ring yields similar velocity profiles for different decay indices $n$ of the 
external poloidal field $B_\mathrm{ex}$ (equilibrium field strengths at
the initial radius $R_0\!=\!10$~Mm are given).\label{fig1}}
\end{minipage}
\end{figure*}

Theoretical models of CMEs did not address the wide range of CME
kinematics for many years. Some modellers reproduced the trajectories
of individual filament eruptions and CMEs in good agreement with the
observations (e.g., Krall et al. 2001; Chen \& Krall 2003; T\"or\"ok \&
Kliem 2004, 2005; Williams et al. 2005; Chen et al. 2006).
A first attempt to model impulsive and gradual CMEs in a unified manner
was presented by Chen \& Krall (2003). These authors enforced the
expansion of a line-tied, semi-toroidal flux rope by driving it out of
equilibrium through the ``injection'' of poloidal magnetic flux at the
photospheric level. The acceleration of the rope is dominated by its
Lorentz self-force (hoop force) and significant only during the period of
flux injection. A range of acceleration profiles of the rope apex,
encompassing impulsive and gradual CMEs, was obtained by prescribing
different amplidudes and durations of the injection and by varying
geometrical parameters of the rope.

Recently, we suggested a flux rope model which explains the wide range of
CME kinematics without the need of such external driving (Kliem \&
T\"or\"ok 2006, hereafter KT06). As in the model of Chen \& Krall (2003),
the flux rope expands due to the hoop force, but the eruption is caused
by an ideal MHD instability, the torus instability (TI). A range of flux
rope trajectories encompassing impulsive and gradual events results
naturally from the model if the height dependence of the magnetic field
overlying the flux rope is varied.
However, the analytical treatment of CME acceleration in KT06 required
simplifications, notably a freely expanding circular current ring (rather
than a line-tied flux rope) was considered, and flux pileup, pressure
gradients, and external toroidal fields had to be neglected.

In this paper, we present MHD simulations of the TI which
include the effects of photospheric line tying and of flux pileup in front
of the expanding rope. By considering a substantial range of magnetic
field profiles above the flux rope, we confirm our previous result (KT06)
that the TI of a flux rope provides a CME model which describes both
impulsive and gradual CMEs and their association with the main magnetic
structural properties of their source region.

%=======================================================================
\section{Torus instability}
%=======================================================================
Observations of erupting filaments, which often evolve into the cores of CMEs, 
suggest the magnetic topology of a single arched magnetic flux rope, or partial 
current ring, whose footpoints are anchored in the solar surface. As is well
known from fusion research, the required pre-eruption equilibrium in a
low-beta environment must include an external
poloidal magnetic field $\vec{B}_\mathrm{ex}$, since the hoop force, as
well as the net pressure gradient
force, of a bent non-neutralized current channel always point radially
outward. Such a current ring is unstable
against expansion if its hoop force decreases more slowly with major ring radius 
$R$ than the opposing Lorentz force due to $B_\mathrm{ex}$ (Bateman 1978).
Assuming $B_\mathrm{ex}\propto R^{-n}$, where $n=-R\,d\,\mathrm{ln}B_\mathrm{ex}/dR$ is the 
decay index of the external field, Bateman (1978) found $n>n_\mathrm{cr}=3/2$ as condition 
for the instability.

KT06 suggested the instability as a possible initiation and
driving mechanism for CMEs, referring to it as torus instability, and
treated its evolution for the first time
(using the same external field as Bateman 1978).
It was found that the acceleration profile of the freely expanding current ring 
strongly depends on the steepness of the external field decrease with
$R$. For moderate to steep decrease, $n \gtrsim 2$, the
acceleration rises quickly to a peak within $R/R_0 \lesssim 2$ and
decreases quickly afterwards, resulting in a profile as observed in
impulsive CMEs. The peak acceleration increases, and its radial position
decreases, with increasing $n$, so that the CME develops a more
impulsive trajectory if it originates in more localized and more
complex flux concentrations. For very gradual decrease ($n$ approaching
$n_{cr}$), the acceleration
has a gradual profile, nearly uniformly distributed
over a large radial range, i.e., with small peak value shifted to
larger $R$; see Fig.\,1 in KT06.

Hence, the instability produces not only a continuum of acceleration
profiles ranging from fast impulsive to slow gradual ones, it also
reproduces the observed association of impulsive profiles with highly
concentrated flux in the source (active regions, high $n$) and of gradual
profiles with the gradual field decrease over a large height range in the
quiet Sun ($n\sim3/2$, Vr{\v s}nak et al. 2002).
Fig.\,1 shows a comparison of velocity profiles obtained from this model
% (for $R_0\!=\!10$~Mm) 
with observed CME velocities.
We note that more gradual profiles result also if $R_0$
is increased (which compresses the $x$
axis of the right panel in Fig.\,1 proportionally); this is the effect used by Chen
\& Krall (2003). 
Both small $n$ and large $R_0$ make
large-scale quiescent filament eruptions gradual.
 
\begin{figure}[t]
\begin{center}
\includegraphics[width=0.84\linewidth]{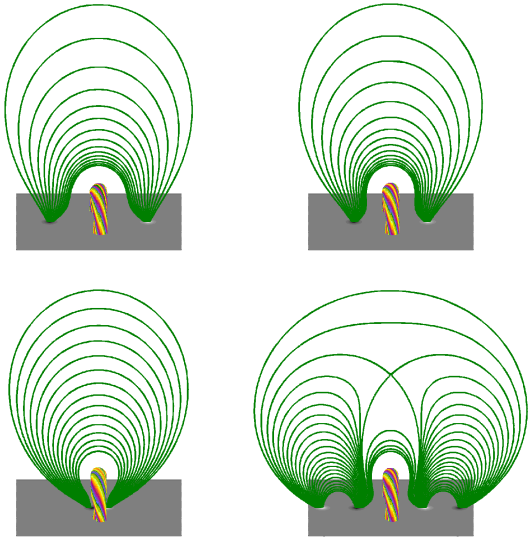}  %90
\end{center}
\caption{Initial configurations of the numerical simulations, corresponding to 
the red, blue, green, and black lines in Figure 3, respectively (from top left 
to bottom right). The rainbow-coloured field lines show the centre of the flux 
rope (current ring), the green field lines show the overlying magnetic field 
(which is a superposition of the external poloidal field $B_{ex}$ and the poloidal 
field created by the ring current).
Normalized distances between the monopoles are $3.6$, $2.7$, 
and $0.5\,h_0$ in the bipolar configurations and $2.5$ and $5.0\,h_0$ 
in the quadrupolar one.}  
\label{fig2}
\end{figure} 

\begin{figure}[t]
\begin{center}
\includegraphics[width=0.76\linewidth]{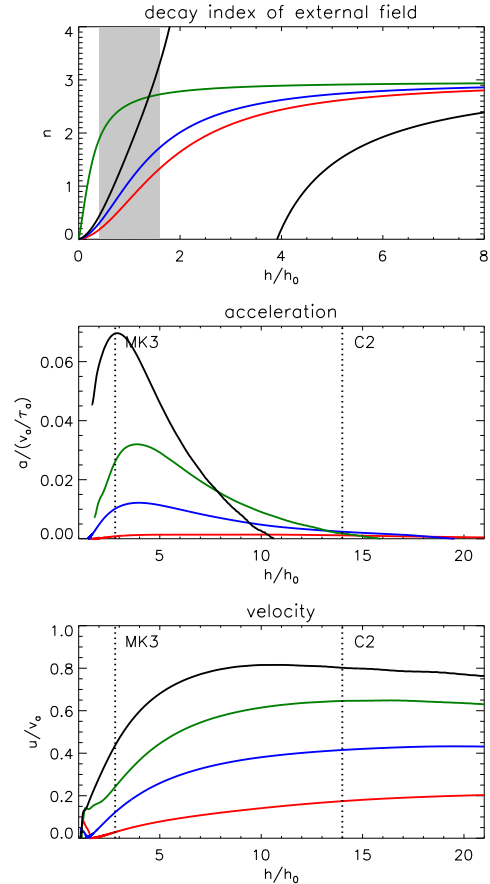}  %78
\end{center}
\caption{Decay indices $n(h)=-h\,d\,\mathrm{ln}B_\mathrm{ex}/dh$ of the external 
poloidal field and 
corresponding acceleration and velocity profiles for the torus-unstable flux ropes 
shown in Fig.\,2 ($|n|\to\infty$ at the null in the quadrupolar configuration). 
The initial position of the current-carrying flux rope is indicated in grey. 
The inner edges of the MK3 and LASCO/C2 coronagraph 
fields of view are indicated for a scaled initial apex height of the flux rope of 
$50$~Mm. Velocities are normalized to the initial Alfv\'en velocity $v_a$ at the
flux rope apex; $\tau_a=h_0/v_a$. The acceleration during the initial relaxation phase 
is omitted for clarity of the figure.}
\label{fig3}
\end{figure}

%=======================================================================
\section{Numerical simulations}
%=======================================================================
In order to confirm the analytical results,
% described in the previous section for a more realistic CME model,
we perform 3D zero-beta ideal MHD simulations of the TI, using the 
approximately force-free coronal flux rope model by Titov \& D\'emoulin (1999) 
as initial condition. The line-tied flux rope corresponds to a current ring partly 
submerged in the dense subphotospheric plasma.
% , hence including the influence of line tying. 

The setup and initial conditions of the simulations are similar to those 
described in T\"or\"ok \& Kliem (2005). Here we use a flux rope with a 
smaller aspect ratio by setting its minor radius to $6/10$ of its initial
apex height $h_0$ (to obtain a configuration with a flux rope twist below the 
threshold of the helical kink instability) and we choose the initial flux 
rope shape to be almost semi-circular by setting the depth of the torus 
center below the photosphere to be only 0.1\,$h_0$
(since the line-tying suppresses the TI before 
a semi-circular shape is reached; Vr{\v s}nak, Ru{\v z}djak \& Rompolt
1991; Chen et al. 2006).
Furthermore, we remove the toroidal component of the external field used 
in T\"or\"ok \& Kliem (2005), to facilitate the comparison with the analytical 
TI model. The resulting purely poloidal external field is created by a pair of 
subphotospheric monopoles 
(see Titov \& D\'emoulin 1999). 
It is more realistic
than the one used in KT06, as it yields a decay index which 
increases monotonically with height above the photosphere, $n=n(h)$.

We fix the geometrical flux rope parameters and consider different 
external field profiles by varying the number, position and strength of the 
subphotospheric monopoles in the model (such that the variations of 
the ring current and the flux rope twist stay below 1\%). 
The resulting configurations are shown in Fig.\,2. 
The field above the flux rope of the configuration in the bottom right panel is 
quadrupolar and contains an X-line. The top panel in Fig.\,3 
shows the corresponding decay indices $n(h)$ of the external fields. The external field 
in the quadrupolar configuration drops very rapidly with height, with $n(h)>3$ 
already in the region immediately above the rope. The other configurations all have 
external fields with $n(h)<3$ above the rope, but with different slopes. 
As $h \rightarrow \infty$, $n \rightarrow 3$ for all configurations.
 
The TI is found to occur also in the numerical model. It is triggered by 
moderate upward motions
set up in the initial relaxation phase of the only approximatively force-free 
configuration, which lift the flux rope to a height where the conditions for TI onset ($2h_0\gtrsim D$,
$D$ being the footpoint distance, and
$n(h)>n_\mathrm{cr}$) are fulfilled. In the case 
shown in the top left panel in Fig.\,2, 
the field decrease is so gradual that the initial relaxation does not move the 
flux rope to the height required for TI onset,
the rope finds a stable numerical equilibrium after relaxation. In order to
trigger the TI in this configuration, we artificially set
the monopole strength to be 9\% smaller than its  
equilibrium value, which causes the rope to relax towards an equilibrium at a
slightly greater height, sufficient for TI onset.
The simulations show agreement with the analytically obtained instability threshold:
in the two runs with the lowest $n(h)$ profile, the TI starts at flux
rope apex heights where $n\lesssim1.5$.

Despite the differences between the analytical and the numerical model, the rise
profiles are qualitatively very similar (compare Fig.\,3 with Fig.\,1 in KT06).
%Again, the acceleration rises quickly to a maximum which strongly 
%increases with the steepness of the external field decrease. Afterwards, the
%steeper the external field decreases, the faster drops the acceleration. 
%The peak acceleration is shifted to somewhat greater heights also for
%fast ejections, $h/h_0\sim2\mbox{--}4$, which appears to conform better
%to the observations than the analytical result ($h/h_0\approx2$).
In both cases, the acceleration rises quickly to a maximum which strongly increases 
with the steepness of the external field decrease. Afterwards, the steeper the field 
decrease, the steeper the drop of the acceleration. In the numerical model, the peak
acceleration is shifted to somewhat greater heights for fast ejections, 
$h/h_0\sim2\mbox{--}4$, which appears to conform better to the observations than the 
analytical result ($h/h_0\approx2$).

The quadrupolar configuration yields the most impulsive acceleration and
the highest terminal velocity.

%=======================================================================
\section{Conclusions and discussion}
%=======================================================================
We present numerical simulations of an erupting force-free magnetic flux 
rope embedded in the potential field of model bipolar and quadrupolar active 
regions. The eruption is driven by the TI which occurs if the external 
poloidal field 
decreases sufficiently rapidly with height. The simulations show that the TI 
occurs not only in freely expanding current rings (KT06), 
but also in line-tied flux ropes. The TI threshold is similar in both cases, 
$n_\mathrm{cr} \sim 3/2$. 
 
The acceleration profile depends on the steepness of the field decrease, 
corresponding to fast CMEs for rapid decrease (as is typical of active 
regions) and to slow CMEs for gradual decrease (as is typical of the quiet 
Sun). These simulations confirm our previous finding (KT06) that 
the TI permits to describe fast and slow CMEs in a unfied manner. 
This demonstrates, in accordance with the observational results in
Vr{\v s}nak, Sudar \& Ru{\v z}djak (2005) and Zhang \& Dere (2006),
% of Chen \& Krall (2003) and with recent observations,
that the concept of two separate CME types is
not valid and that the wide range of CME kinematics can be explained by
a single underlying physical process.

The ratio of the maximum velocities (accelerations) in the fastest and
the  slowest eruption in Fig.~\ref{fig3} is $\approx 4$ ($\approx 40$). These ranges are about
one order of magnitude smaller than the corresponding ranges observed in
CMEs. However, it is important to note that the peak velocities and
accelerations listed in the Introduction are matched with very reasonable
choices for the Alfv\'en speed in the core of large active regions
($v_a\sim3750$~km\,s$^{-1}$) and for $h_0$ ($\lesssim R_\odot/10$),
although the inclusion of line tying and flux pileup in
front of the expanding rope in the numerical model reduce the
acceleration considerably in comparison to the analytical result.
The great majority of observed CME
speeds and accelerations is accommodated for if the variations of $v_a$
and $h_0$ between $\delta$ spots and quiescent filaments, both by factors
3--10, are taken into account. It is always possible to match extremely
slow CMEs by setting the parameters of the simulation very close to the
instability threshold. Moreover, for gradual CMEs which result from
eruptions of quiescent filaments and are accelerated over several
$R_\odot$, the presence of the heliospheric current sheet reduces the
decay index of the overlying field in comparison to the asymptotic value
($n=3$) in the Titov \& D\'emoulin equilibrium.

Magnetic fields in multipolar active regions decrease in the lower 
corona faster with height than those in bipolar active regions or in the quiet 
Sun. Consequently, our model predicts the fastest CMEs to 
originate in multipolar active regions, which is in line with 
observations. No ``breakout-like'' reconnection above the flux rope is 
required in our simulation of the quadrupolar model configuration, the
eruption is solely driven by the TI.
 
In contrast to the model of Chen \& Krall (2003), our model allows to 
obtain a wide range of CME kinematics without imposing an external driving of the 
configuration with a duration similar to the observed duration of
significant acceleration. For given flux rope parameters, the kinematics 
are determined by the slope of the magnetic field 
overlying the flux rope. Although the magnetic field in the corona 
cannot be measured directly at present, a wide range of  
height profiles of the field can be expected to exist, 
as the large variety of field distributions revealed by 
photospheric magnetograms indicates.  

We note that, for a given overlying field, the acceleration profile also depends 
on the chosen flux rope parameters as well as on the existence of a
flux rope velocity at TI onset
(due to external perturbations; Schrijver et al. 2007). 
Furthermore, we expect that an external toroidal field and pressure gradients will
influence the evolution as well.
These effects will be
studied in forthcoming work.

\acknowledgements
We thank the referee for constructive comments.
T. T. thanks the British Council and PPARC for financial support. 
B. K. was supported by the DFG.

\end{document}